\documentclass[prb,twocolumn]{revtex4}

\usepackage{graphicx}

\begin{document}

\author{O. Leenaerts}
\email{ortwin.leenaerts@ua.ac.be} \affiliation{Universiteit Antwerpen, Departement Fysica,
Groenenborgerlaan 171, B-2020 Antwerpen, Belgium}
\author{B. Partoens}
\email{bart.partoens@ua.ac.be} \affiliation{Universiteit Antwerpen, Departement Fysica,
Groenenborgerlaan 171, B-2020 Antwerpen, Belgium}
\author{F. M. Peeters}
\email{francois.peeters@ua.ac.be} \affiliation{Universiteit Antwerpen, Departement Fysica,
Groenenborgerlaan 171, B-2020 Antwerpen, Belgium}
\date{\today}
\title{Graphene: a perfect nanoballoon}

\begin{abstract}
We have performed a first-principles density functional theory investigation of the penetration of helium atoms through a graphene monolayer with defects. The relaxation of the graphene layer caused by the incoming helium atoms does not have a strong influence on the height of the energy barriers for penetration. For defective graphene layers, the penetration barriers decrease exponentially with the size of the defects but they are still sufficiently high that very large defects are needed to make the graphene sheet permeable for small atoms and molecules. This makes graphene a very promising material for the construction of nanocages and nanomembranes.
\end{abstract}

\maketitle

Graphene is one of the most studied materials these days, which has resulted already in a large amount of proposals for possible applications.\cite{geim} These applications range from very sensitive gas sensors\cite{schedin} to carbon-based electronics\cite{avouris} and are mainly based on the essentially two-dimensional (2D) form of graphene and the Dirac-like behavior of the electrons at the Fermi level.\cite{katsnelson} Recently it was experimentally shown that perfect graphene sheets are impermeable to standard gases, including helium.\cite{bunch} This introduces a new range of applications for graphene as an ultrathin, but still impermeable, membrane. In Ref. \onlinecite{bunch}, Bunch et al. suggested also that the graphene samples should be free of defects to explain the impermeability. This suggestion was based on a simple classical effusion theory calculation of the penetration of point particles through single atom vacancies in graphene. 
In this letter, however, we demonstrate through {\it ab initio} calculations that defective graphene is still impermeable and that large defects are needed to destroy this impermeability. In our study we concentrate on the penetration of helium atoms through graphene with increasingly large defects. Helium atoms are the smallest atoms that do not chemically interact with graphene. We limit ourselves to point defects that keep the sp$^{2}$ hybridization of the carbon atoms of graphene more or less intact. Such defects are more stable\cite{lee1,lee2} and easier to treat in first-principles calculations. 

We make use of the density functional theory (DFT) formalism in both the local density (LDA) and general gradient approximation (GGA). All our DFT calculations are performed with the ABINIT\cite{abinit} software package. The simulation of most defects is done in a $4\times 4\times 4$ graphene supercell with a distance of 16 {\AA} between adjacent graphene layers. A plane-wave basis with a cutoff energy of 30 Hartree (816 eV) was used and the Brioullin zone (BZ) is sampled with a $6\times 6\times 6$ Monckhorst-Pack (MP)\cite{mp} k-point grid which is equivalent to a $24\times 24\times 24$ MP grid in a single unit cell. We used pseudopotentials of the Troullier-Martins type\cite{tm} for both the LDA and GGA calculations. There is no need to perform spin-polarized calculations because the defects were chosen to preserve the sp$^{2}$ hybridization of the carbon atoms in the simulated defective graphene sheets and the He atom is inert.

\begin{figure}[h]
  \centering
\includegraphics[width= 3.25 in]{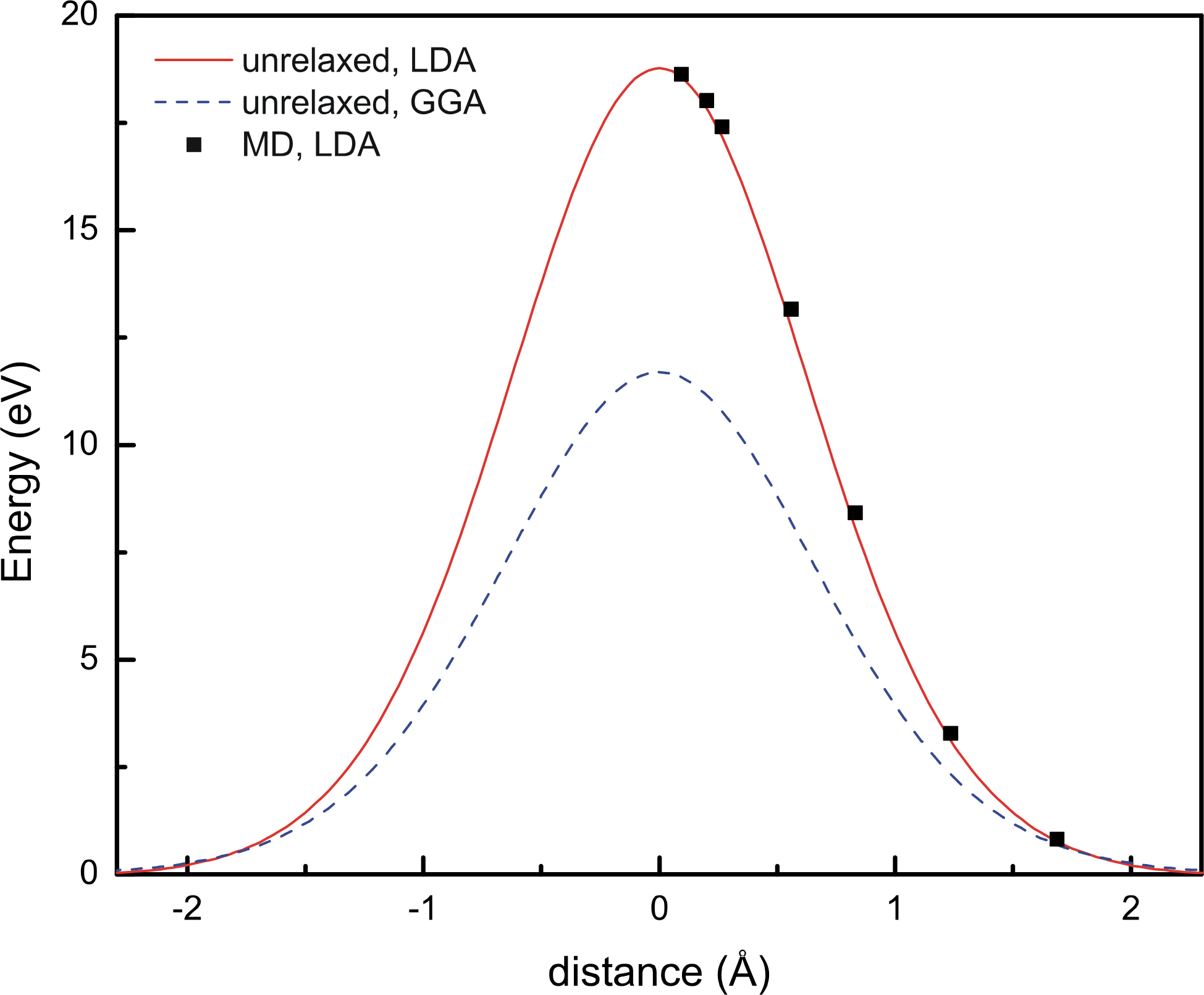}
\caption{\label{fig1}(Color online) The potential barrier for the penetration of a He atom through the center of a carbon hexagon of a perfect graphene layer. The results for the penetration without relaxation of the graphene layer are shown for LDA and GGA (solid and dashed line respectively). The results of the MD calculations (LDA) are indicated by full squares.}
\end{figure}

We first examine the penetration of a helium atom through the center of a carbon hexagon of a perfect graphene monolayer. As a first approximation, we keep all the carbon atoms fixed and calculate the potential energy of the He atom at different distances from the graphene sheet. The resulting energy barrier is given in Fig. \ref{fig1} for both LDA and GGA. The height of the barrier, 18.8 eV for LDA and 11.7 for GGA, is very large and makes penetration of helium gas that is in thermal equilibrium impossible at any temperature at which the graphene layer remains stable. 

\begin{figure}[h]
  \centering
\includegraphics[width= 3.25 in]{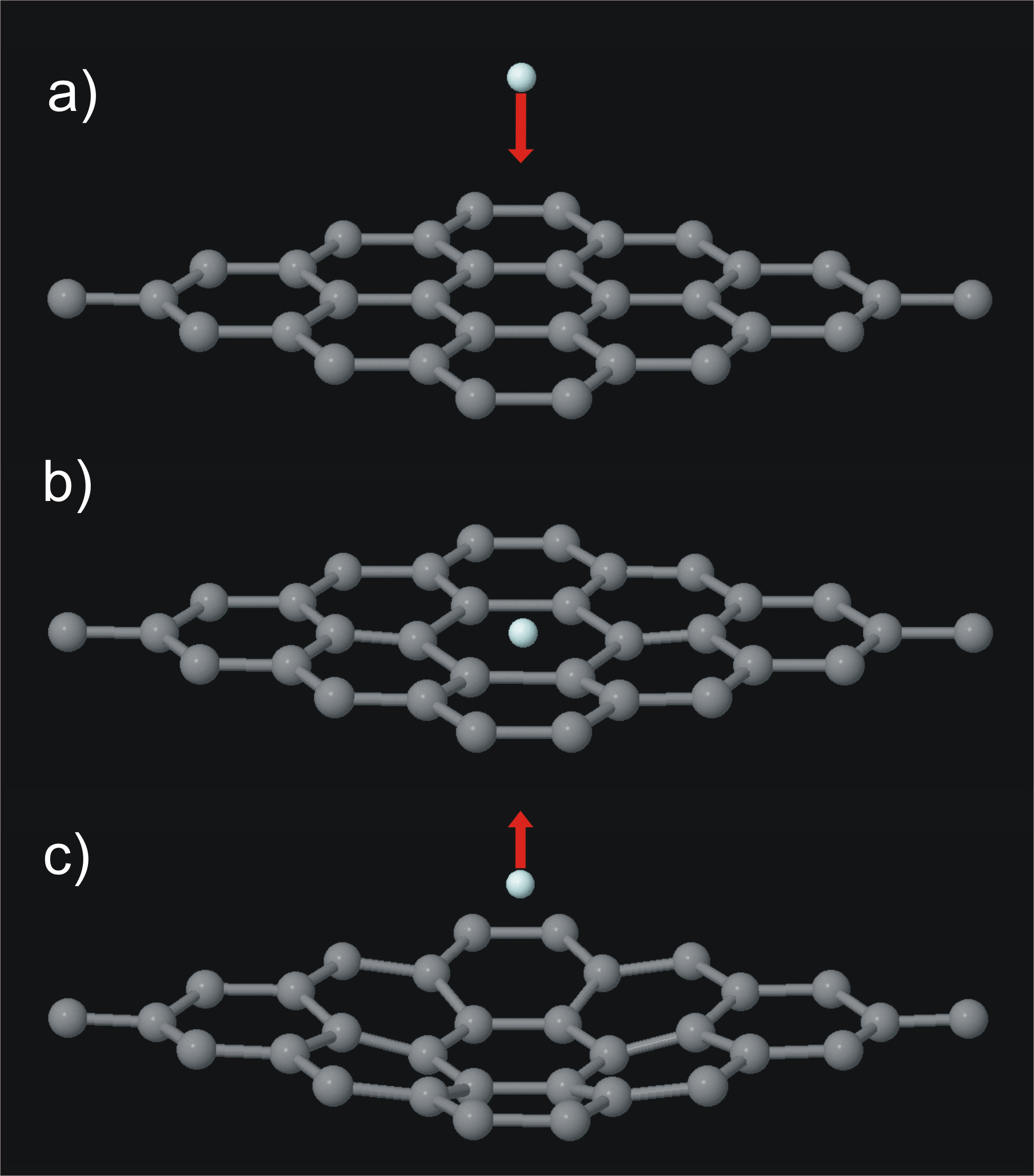}
\caption{\label{fig2}(Color online) Reflection of a He atom with a kinetic energy of 18.6 eV from a graphene surface: a) The He atom approaches the perfect graphene layer. b) The He atom comes to rest before penetrating the graphene layer. Note that the relaxation of the graphene layer is very small at this moment. c) The He atom is reflected back and the surface starts to relax.}
\end{figure}

Next we used velocity-Verlet molecular dynamics\cite{verlet1,verlet2} simulations to show that the height of the barrier is not essentially decreased when we allow the graphene layer to relax when the helium atom impinges on the layer. These MD calculations are done within the DFT formalism (LDA) and with full relaxation of the graphene sheet. We start with a nonperturbed graphene sheet and place the He atom above the center of a graphene hexagon at a distance far from the graphene surface where the potential energy is vanishingly small. Then we give the He atom different velocities in the direction of the graphene layer. When the kinetic energy of the He atom is small enough, it will bounce back from the graphene surface at a certain distance d, which we define to be the difference along the $z$ direction (perpendicular to the graphene surface) between the He atom and the closest C atom of the graphene sheet. The effective potential felt by the He atom depends on its velocity due to the relaxation of the graphene layer. This potential can be constructed from our MD calculations by considering the kinetic energy of the incoming He atom as a function of the distance d at which it is reflected from the graphene layer.
The results of the MD calculations are shown in Fig. \ref{fig1} by the black squares. Notice that the difference of the potential felt by the approaching He atom between the relaxed (MD) and the unrelaxed case is very small. This is a consequence of the fact that the graphene atoms appear to lack time to relax while interacting with the fast moving He atom. In Fig. \ref{fig2} the reflection of a He atom with a kinetic energy that is just a little smaller that the energy barrier for penetration is shown. It is clear that the relaxation of the graphene layer is very small at the turning of the He atom (see Fig. \ref{fig2}b) and that the relaxation only starts when the He has already been reflected (see Fig. \ref{fig2}c). This means that relaxation has no significant influence on the barrier height and from now on we are allowed to ignore any relaxation of the graphene layer when calculating the energy barriers for penetration of the He atom.

Our approach is different from earlier calculations of the barrier height for the penetration of He into C$_{60}$ fullerenes,\cite{hrusak} where the He atom was given a fixed position and the system was allowed to relax completely before calculating the barrier. In our opinion this is actually a less realistic calculation than one where any relaxation is ignored.  The reason is that the system relaxes within a much larger time scale than the time scale for motion of the much lighter, chemically inert, He atom. Note that this is in general not always true: in the case of other atoms, e.g. a hydrogen atom,\cite{ito} or molecules, the interaction time can be long enough for the graphene layer to have a significant relaxation, which may be due to the much larger mass of the molecules or as a consequence of chemical reactivity.

\begin{figure}[h]
  \centering
\includegraphics[width= 3.25 in]{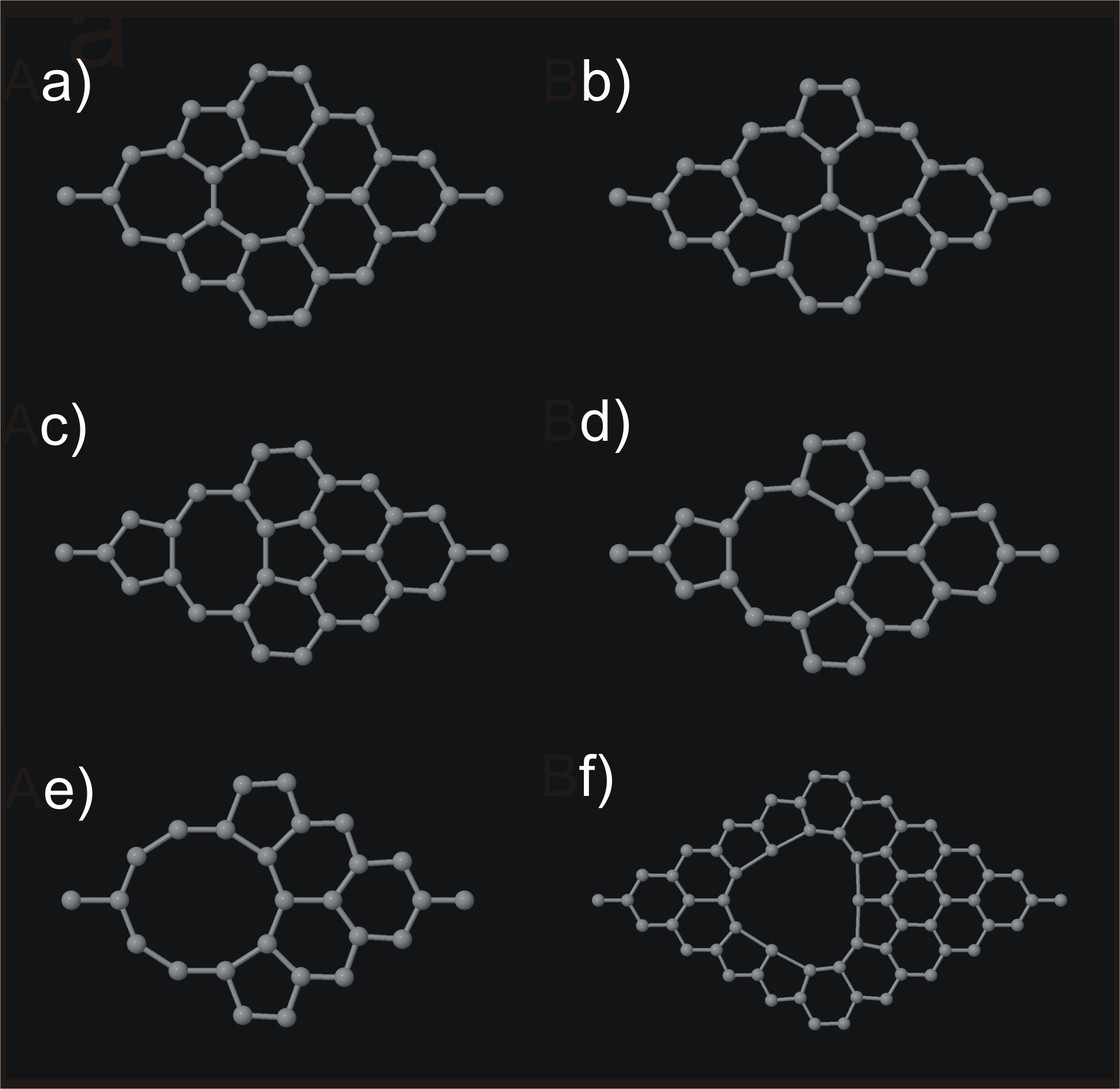}
\caption{\label{fig3}(Color online) The considered sp$^{2}$ hybridization conserving defects: a) Stone-Wales defect, b) 555-777 divacancy, c) 585 divacancy, d) tetravacancy, e) hexavacancy, and f) decavacancy.}
\end{figure}

We now turn our attention to the penetration of a defective graphene layer where we consider only those defects that do not destroy the sp$^{2}$ hybridization of the carbon atoms. Examples of studied defects (see Fig. \ref{fig3}) include the Stone-Wales (SW) defect, the divacancy (585 and 555777), and the tetra-, hexa- and decavacancy. All these defects were fully relaxed in a $4\times 4\times 4$ graphene supercell, but for the decavacancy a $6\times 6\times 6$ supercell was used. The lowest penetration barrier is always found inside the largest 'ring' of carbon atoms that belongs to the defect. The resulting barriers for the different defects are given in Table \ref{tab1}. 

\begin{table}[h]
\caption{The energy barrier height (in eV) for penetration of a He atom through perfect and defective graphene as obtained within LDA and GGA.\label{tab1}}
\begin{tabular}{lcc}
\hline\hline
{\bf defect}     & {\bf LDA}  &	  {\bf GGA}     \\
\hline

       no defect   			&  18.77 & 11.69   \\

       Stone-Wales 			&  9.21 &  6.12   \\

       555777 divacancy &  8.77 &  5.75   \\

       858 divacancy    &  4.61 &  3.35   \\
       
       tetravacancy  		&  1.20 &  1.04   \\
       
       hexavacancy  		&  0.37 &  0.44   \\
       
       decavacancy  		&  0.05 &  0.10   \\

\hline\hline
\end{tabular}
\end{table}

\begin{figure}[h]
  \centering
\includegraphics[width= 3.25 in]{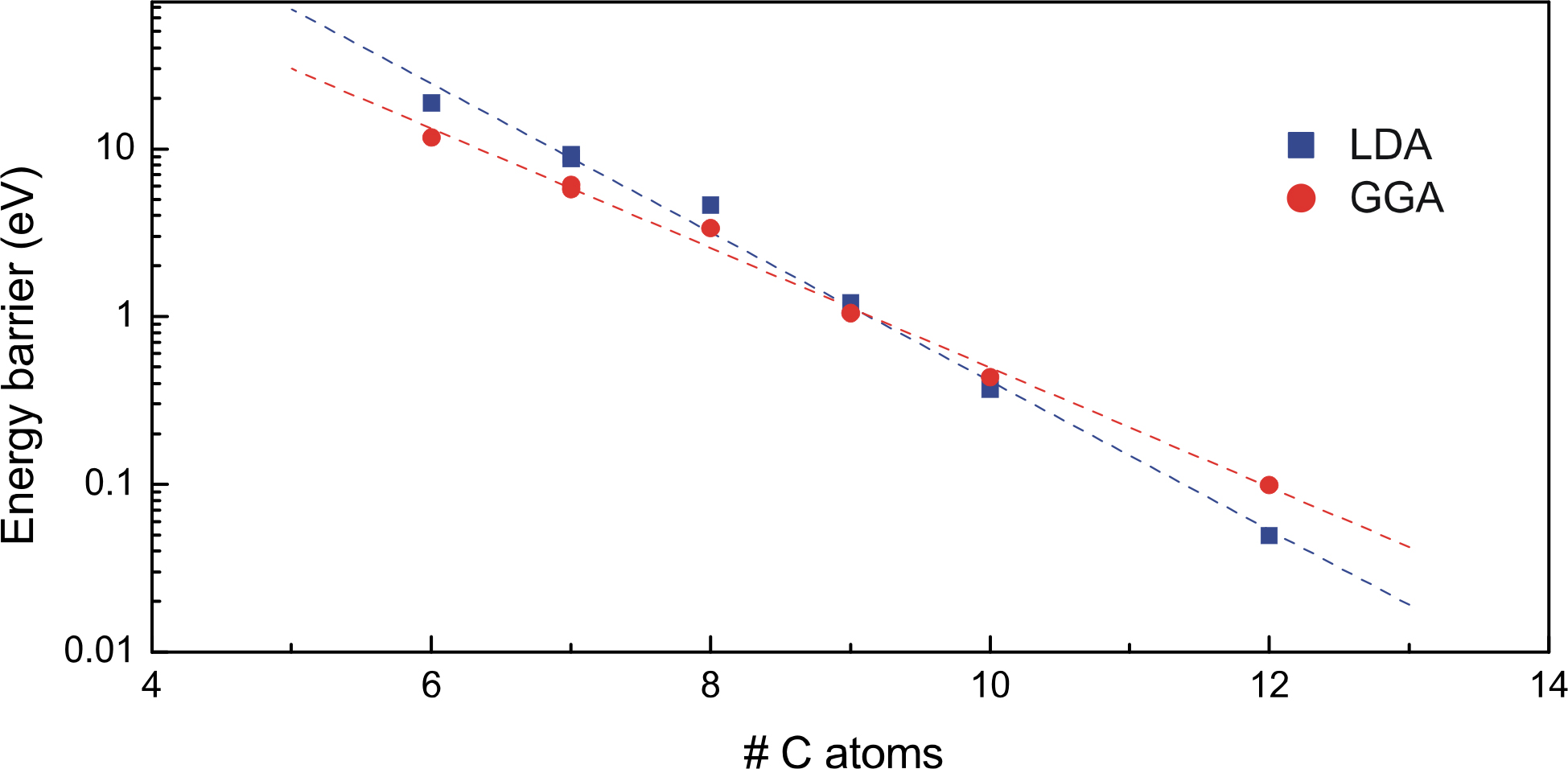}
\caption{\label{fig4}(Color online) The dependence of the penetration barrier height on the size of the defect for LDA and GGA.}
\end{figure}

As intuitively expected, the barrier height decreases fast with increasing size of the defects. A more quantitative relationship between the barrier height and the size of the defect can be obtained by quantifying the defect size as the number of carbon atoms included in the sp$^{2}$-bonded defect ring. In this way, we obtain sizes ranging from 6 (for perfect graphene) to 12 (for the decavacancy). The Stone-Wales defect and the 555777 divacancy both have a size of 7. The barrier height versus defect size is shown in Fig. \ref{fig4}. Notice that, to a very good approximation, the penetration barrier height decreases exponentially with the size of the defect (as indicated by the dashed lines). The largest deviation from the exponential behavior is found for the 585 divacancy with 8 carbon atoms in the defect ring. This is clearly a consequence of the noncircular form of the defect (see Fig. \ref{fig3}c) which is more pronounced in this defect than in the others. The larger the deviation of the circular form of the defect, the worse is the quantification of the defect size through the number of carbon atoms in the defect ring. 

It is also clear from Fig. \ref{fig4} that, although the energy barrier for penetration decreases exponentially, large defects are needed to allow for any appreciable gas leak through a graphene membrane at room temperature ($k_{B}T$ =26 meV). Our findings are in marked contrast to the claim of Ref. \onlinecite{bunch} that it is necessary to have a perfect graphene layer to preserve the impermeability. This claim was based on a simple classical effusion theory calculation which is clearly a nonrealistic approximation for the penetration process of He atoms through defective graphene membranes. Our first-principles calculations clearly suggest that small point defects in graphene will not destroy the impermeability of the membrane.

In summary, we investigated the penetration of a helium atom through perfect and defective graphene. We found that the relaxation of the graphene layer caused by the impact of the incoming He atom has no influence on the barrier height and therefore can be neglected. This is due to the fact that the time scale for relaxation of the graphene is larger than the time scale during the He atom interacts with the graphene layer. The relaxation of the graphene layer occurs after the He atom has left the graphene layer. We found that the penetration barrier height decreases exponentially with the size of the defects, as expressed by the number of carbon atoms included in the formation of the defect. But the penetration barrier height of small defects is still large enough to preserve the impermeability of graphene membranes for atoms and molecules. Consequently, even defective graphene is a suitable candidate for making impermeable nanomembranes for future applications and therefore can be considered the thinnest possible material for constructing a micro- or nanoballoon.

\begin{acknowledgments}
This work was supported by the Flemish Science Foundation (FWO-Vl), the NOI-BOF of the University
of Antwerp and the Belgian Science Policy (IAP). 
\end{acknowledgments}


\begin{thebibliography}{99}


\bibitem{geim} A. K. Geim, and K. S. Novoselov,  Nat. Mater. {\bf 6}(3), 183 (2007).

\bibitem{schedin} F. Schedin, A. K. Geim, S. V. Morozov, E. W. Hill, P. Blake, M. I. Katsnelson, and K. S. Novosolov,  Nat. Mat. {\bf 6}, 652 (2007).

\bibitem{avouris} P.	Avouris, Z. Chen, and V. Perebeinos,  Nat.  Nanothech. {\bf 2}, 605 (2007).

\bibitem{katsnelson} M. I.	Katsnelson,  Mat. Tod. {\bf 10}, 20 (2007).

\bibitem{bunch}	J. S. Bunch, S. S. Verbridge, J. S. Alden, A. M. van der Zande, J. M. Parpia, H. G. Craighead, and P. L. McEuen, Nano Lett. {\bf 8}(8), 2458 (2008).

\bibitem{lee1} G. D.	Lee, C. Z. Wang, E. Yoon, N. M. Hwang, D. Y. Kim, and K. M. Ho,  Phys. Rev. Lett. {\bf 95}, 205501 (2005).

\bibitem{lee2} G. D.	Lee, C. Z. Wang, E. Yoon, N. M. Hwang, D. Y. Kim, and K. M. Ho,  Phys. Rev. B {\bf 74}, 245411 (2006).

\bibitem{abinit}	URL: http://www.abinit.org/

\bibitem{mp} H. J.	Monkhorst, and J. D. Pack,  Phys. Rev. B {\bf 13}, 5188 (1971).

\bibitem{tm}	N. Troullier, and J. L. Martins, Phys. Rev. B {\bf 43}, 1993 (1991).

\bibitem{verlet1} L. 	Verlet, Phys. Rev. {\bf 159},98 (1967).

\bibitem{verlet2} L.	Verlet, Phys. Rev. {\bf 165}, 201 (1967).

\bibitem{hrusak} J.	Hrušák, D. K. Böhme, T. Weiske, and H. Schwarz,  Chem. Phys. Lett. {\bf 193}, 97 (1992).

\bibitem{ito}	A. Ito, H. Nakamura, and A. Takayama,  cond-mat/0703377.


\end{thebibliography}
\end{document}